\documentclass[a4paper,10pt]{article}
\usepackage{amsmath}
\usepackage{amssymb}
\usepackage{array}
\usepackage{calc}
\usepackage{longtable}
\usepackage{multirow,booktabs}
\usepackage{cite,mcite}
\usepackage{relsize}
\usepackage{pstricks}
\usepackage{graphicx}
\usepackage{xspace}
\usepackage{units}
\usepackage{afterpage}
\usepackage{placeins}
\usepackage[compat=1.1.0]{tikz-feynman}

\numberwithin{equation}{section}
\usepackage{mciteplus}
\usepackage[pdfborder={0 0 0}]{hyperref}
\usepackage[format=hang,labelfont=bf,hypcap=true]{caption}
\usepackage{subcaption}
\usepackage{sectsty}
\allsectionsfont{\sffamily}
\subsubsectionfont{\mdseries\itshape\large}
\makeatletter
\DeclareRobustCommand*{\bfseries}{%
  \not@math@alphabet\bfseries\mathbf
  \fontseries\bfdefault\selectfont
  \boldmath
}
\makeatother
\let\spreprint\empty
\newcommand{\preprint}[1]{\def\spreprint{\protect#1}}
\let\sinstitute\empty
\newcommand{\institute}[1]{\def\sinstitute{\protect#1}}
\makeatletter
\renewcommand{\maketitle}{\begingroup
  \null\thispagestyle{empty}%
    \ifx\spreprint\empty
      \vskip 5ex
    \else
      \flushright\large\spreprint\vskip 2ex
    \fi
    \vskip 5ex
    \flushleft
      {\sffamily\bfseries\huge\@title}\vskip 6ex
      \@author\vskip 2ex
      \ifx\sinstitute\empty
      \else
        {\small\sinstitute}
      \fi
    \vskip 5ex
  \endgroup
}
\makeatother
\renewenvironment{abstract}{\begin{center}
  {\large\sffamily\bfseries Abstract: }
  \begin{minipage}[t]{0.75\textwidth}
}{\end{minipage}\end{center}\vskip 10ex}

\numberwithin{equation}{section}

\newcommand{\bea}{\begin{eqnarray}}
\newcommand{\eea}{\end{eqnarray}}
\newcommand{\beq}{\begin{equation}}
\newcommand{\eeq}{\end{equation}}

\newcommand{\bs}{\begin{split}}
\newcommand{\es}{\end{split}}

\newcommand{\bi}{\begin{itemize}}
\newcommand{\ei}{\end{itemize}}
\newcommand{\bc}{\begin{center}}
\newcommand{\ec}{\end{center}}
\newcommand{\bac}{\begin{array}{c}}
\newcommand{\bacc}{\begin{array}{cc}}
\newcommand{\ea}{\end{array}}

\def\spa#1.#2{\langle#1\,#2\rangle}
\def\spb#1.#2{[#1\,#2]}

\newcommand{\sla}[1]{\ensuremath{{#1\kern-0.45em/}}}

\newcommand\DESY{D\scalebox{0.8}{ESY}\xspace}

\newcommand\HERA{H\scalebox{0.8}{ERA}\xspace}
\newcommand\EIC{E\scalebox{0.8}{IC}\xspace}

\newcommand\LHC{L\protect\scalebox{0.8}{HC}\xspace}

\newcommand{\MCatNLO}{M\protect\scalebox{0.8}{C}@N\protect\scalebox{0.8}{LO}\xspace}

\newcommand{\OpenLoops}{O\protect\scalebox{0.8}{PEN}L\protect\scalebox{0.8}{OOPS}\xspace}

\newcommand{\Sherpa}{S\protect\scalebox{0.8}{HERPA}\xspace}
\newcommand{\Comix}{C\protect\scalebox{0.8}{OMIX}\xspace}

\newcommand{\Amegic}{A\protect\scalebox{0.8}{MEGIC}\xspace}
\newcommand{\CSS}{CSS\protect\scalebox{0.8}{HOWER}\xspace}

\newcommand{\Rivet}{R\protect\scalebox{0.8}{IVET}\xspace}

\newcommand{\LHAPDF}{L\protect\scalebox{0.8}{HAPDF}\xspace}

\hypersetup{
  pdfauthor={Peter Meinzinger, Frank Krauss}
  pdftitle={Fully differential NLO predictions for jet photoproduction at the Electron-Ion collider},
  pdfkeywords={QCD, Event Generators, Photon PDF, HERA, LEP, EIC}
}
\begin{document}
\preprint{
  IPPP/23/70}
\title{Hadron-Level NLO predictions for QCD Observables in Photo-Production at the Electron-Ion Collider}
\author{Peter~Meinzinger\footnote[1]{peter.meinzinger@durham.ac.uk}, Frank~Krauss\footnote[2]{frank.krauss@durham.ac.uk}}
\institute{
  Institute for Particle Physics Phenomenology,
  Durham University, Durham DH1 3LE, UK
}

\maketitle
\begin{abstract}
    We present the first next-to leading order accurate hadron-level predictions for inclusive QCD and jet observables in photo-production for the upcoming electron--ion collider. 
\end{abstract}

\section{Introduction}\label{Sec:Intro}
The upcoming electron-ion collider (\EIC) aims at precision measurements of a variety of phenomena related to strong interactions and the impact of a nuclear environment on them. 
The scientific programme~\cite{AbdulKhalek:2021gbh} includes determinations of the strong coupling constant and its scaling behaviour, observations of saturation effects, and measurements of the three-dimensional structure of nucleons as encoded in parton distribution functions (PDFs), generalised parton distributions (GPDs), or transverse-momentum dependent parton disrtibutins functions (TMDs). 
This ambitious programme rests, among others, on high-quality theory predictions, both in the form of analytic calculations and of Monte Carlo simulations, also known as event generators.

Some of the first analyses at the \EIC will investigate the production of inclusive hadronic final states; their properties are usually described through a combination of event shape and jet observables, particle multiplicities and distributions. 
From previous experiments, for example at the \HERA collider at \DESY, it has become customary to classify scattering events by the virtual mass squared, or virtuality, $Q^2$, of the photon exchanged between the incident electron and nucleon (or, in the case of the \EIC, nucleus).
Broadly speaking, essential parts of the \HERA scientific programme, namely the determination of proton structure functions and PDFs, were nearly exclusively driven by the regime of non-zero $Q^2$ associated with deeply-inelastic scattering or electro-production.
The same will certainly be also true for \EIC-based efforts of improved PDF determination.

However, as the spectrum of the exchanged photons scales like $1/Q^2$ it is clear that the electron-\-proton cross section is dominated by the photo-production regime of relatively low $Q^2\approx 0$.
This necessitates the careful treatment of such photo-production events, which so far have not attracted the same attention as, for example, event generation for the \LHC, which has reached a very satisfying level of theoretical accuracy and maturity~\cite{Buckley:2011ms}.
This is due to the fact that in this regime the photon has a non-negligible hadronic component, essentially fluctuating into states with quantum numbers of neutral vector mesons such as the $\rho^0$ or similar. 
At higher resolution scales, this non-perturbative component of the photon structure is further augmented by perturbative splittings of the photon into quark--\-anti-quark pairs -- a feature that is not present in the more familiar proton PDFs.
As a consequence, the corresponding $\gamma p$ collisions will not be characterised by a point-like photon interacting with partons, but rather look like hadron-hadron collisions involving the PDFs of photons, with parametrizations usually dating back two decades.
This in itself suggests an interesting physics programme related to photo-production, evidenced by the breadth and number of published analysis by the \HERA collaborations, which cover a wide range of exclusive and inclusive final states. 

In this paper we build on previous work~\cite{Hoeche:2023gme} and apply the \Sherpa framework for the simulation of hadron-level photo-production events at NLO accuracy to the \EIC.  
The quality of \Sherpa's agreement with \HERA data in this publication gives us confidence that our results presented here will provide a first essential, high-quality baseline for future studies of inclusive QCD final states at this experiment, based on the well-established framework of collinear factorisation.
The code underpinning our work will be made publicly available in a forthcoming release of \Sherpa.

\section{Event Generation}\label{Sec:MC}
In contrast to high-$Q^2$ scattering photo-production is characterised by vanishing photon virtualities of $Q^2\approx 0$.
For the simulation of such events, two different processes have to be considered: the photon can either enter the hard-processes directly, $\gamma j \to j j$, the "direct" component, or it can first undergo a transition to its hadronic component or split into a quark--\-anti-quark pair, followed by the usual QCD scale evolution, leading to the process $j j \to j j$, the so-called "resolved" component. 
The total cross-section therefore deomposes as
\begin{equation}
    \sigma_\text{tot} = \sigma_{\gamma j \to X} + \sigma_{jj \to X}\,,
\end{equation}
where $j$ denotes partons coming from either the proton or the photon.  
Our results, presented below, are obtianed by running \Sherpa, where in each run we generated 10 million events each at \MCatNLO accuracy and at Leading Order for comparison. 
We used the implementation of the photon flux and the photon PDFs as described in~\cite{Hoeche:2023gme}.
Matrix elements were calculated with \Amegic~\cite{Krauss:2001iv} and \Comix~\cite{Gleisberg:2008fv,Gleisberg:2007md} for tree-level and \OpenLoops~\cite{Buccioni:2019sur} for one-loop order. 
QCD Bremsstrahlung was modelled by the \CSS parton shower~\cite{Schumann:2007mg} and matched to the NLO matrix elements with the \Sherpa realization~\cite{Hoeche:2011fd} of the \MCatNLO method~\cite{Frixione:2002ik, Frixione:2003ei}. 
Multiple-parton interactions have been added through an implementation of the Sjostrand-van Zijl model~\cite{Sjostrand:1987su,Schuler:1993wr} and the partons were hadronized with the cluster fragmentation of~\cite{Chahal:2022rid}.
Analogously to~\cite{Hoeche:2023gme}, we used the SAS1M and SAS2M sets as photon PDFs and averaged over these two sets as suggested by their authors~\cite{Schuler1995,Schuler1996a}.
For the proton we used the PDF4LHC21\_40\_pdfas set~\cite{PDF4LHCWorkingGroup:2022cjn} interfaced through \LHAPDF~\cite{Buckley:2014ana} and set the value and running of $\alpha_S$ accordingly. 
For the scales we chose $\mu_F = \mu_R = \mbox{\rm max}\{H_T/2,\,4\,\mathrm{GeV}\}$, and estimated the associated uncertainties with a 7-point scale variation. 
We used \Rivet~\cite{Bierlich:2019rhm} for the event analysis. 

The beam setup was chosen to be an electron and a proton beam with energies of 18 and 275 GeV, respectively, in accordance with the highest energy scenario at the planned \EIC facilities~\cite{AbdulKhalek:2021gbh}. 
Particles were analysed using the anti-$k_T$ jet algorithm with $R=1.0$ and a minimum transverse energy of $E_T>6$ GeV and a pseudo-rapidity $|\eta | < 4$, demanding one jet in the event. 

\section{Predictions for dijet production}\label{Sec:Results}

\begin{figure}[!ht]
    \centering
    \begin{tabular}{cc}
        \includegraphics[width=.4\linewidth]{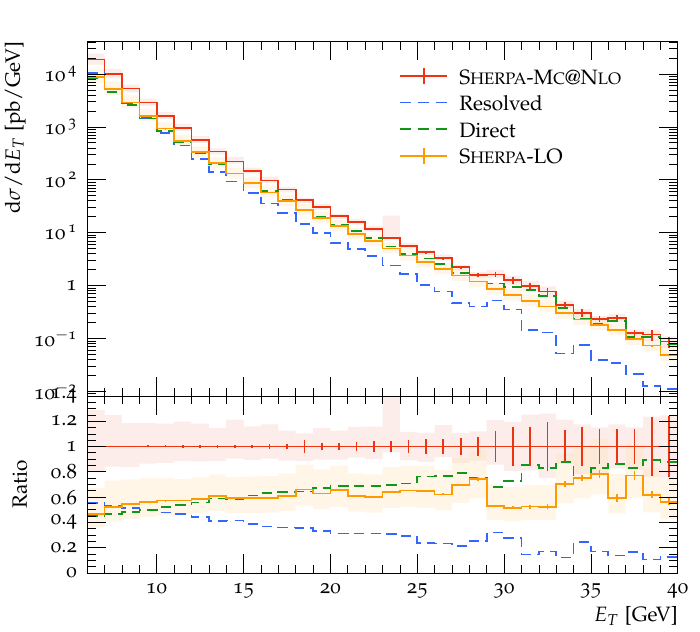} &
        \includegraphics[width=.4\linewidth]{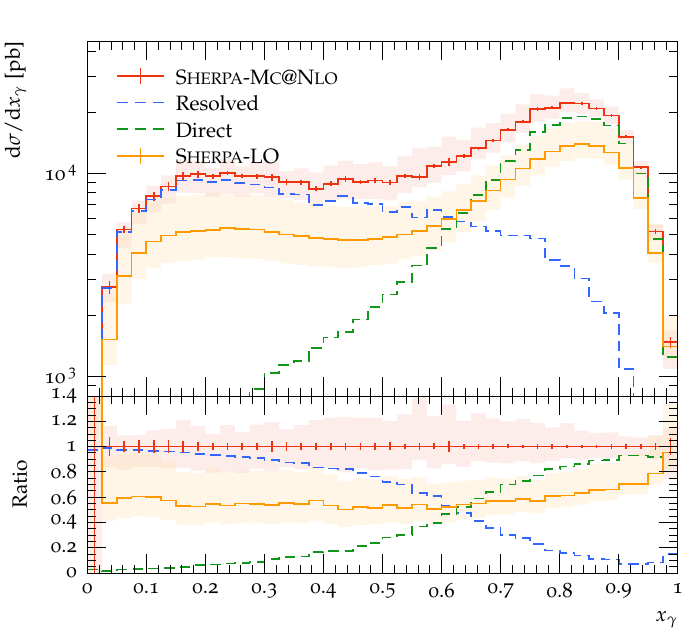}
    \end{tabular}
    \caption{Distributions of inclusive jet transverse energy $E_T$ (left) and the $x_\gamma$ (right) of the \protect\Sherpa simulation with \protect\MCatNLO accuracy, compared with results at LO.}
    \label{fig:et_xgamma}
\end{figure}
In Fig.\ ~\ref{fig:et_xgamma}, we present the inclusive jet transverse momentum distribution and the observable $x_\gamma$, which is defined through
\begin{equation}
    x_\gamma = \frac{E_T^{(1)} \mathrm{e}^{-\eta^{(1)}}+E_T^{(2)} \mathrm{e}^{-\eta^{(2)}}}{2 y E_e}\,.
\end{equation}
Here, $E_T^{(1,2)}$ and $\eta^{(1,2)}$ denote the jet transverse energy and pseudo-rapidity of the leading and sub-leading jet, respectively, and $y$ denotes the fraction of the electron energy $E_e$ carried by the photon. 
It had previously been observed that the $x_\gamma$ observable is especially suitable to discern different photon PDF parametrizations~\cite{ZEUS:2001zoq}. 

The spectrum of the jet transverse energies is driven by the resolved component at $E_T<10$ GeV, with the direct component taking over with increasing jet $E_T$. 
In the $x_\gamma$-distribution the two components are more clearly separated: for low values of $x_\gamma\stackrel{<}{\sim}0,6$ the resolved component dominates, however at $x_\gamma>0.6$ the cross-section of the direct component exceeds the resolved one. 
This is significantly lower than the value of $x_\gamma>0.75$ that had been used as a cut at \HERA~\cite{ZEUS:2001zoq,ZEUS:2012pcn} and is a consequence of the lower collision energy, decreasing the contribution of the resolved component relative to the direct cross-section. 

\begin{figure}[!ht]
    \centering
    \begin{tabular}{cc}
        \includegraphics[width=.4\linewidth]{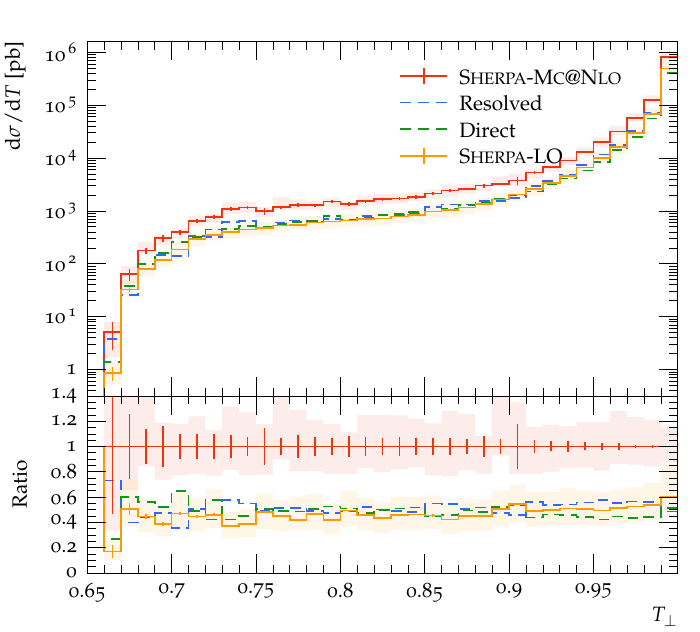} &
        \includegraphics[width=.4\linewidth]{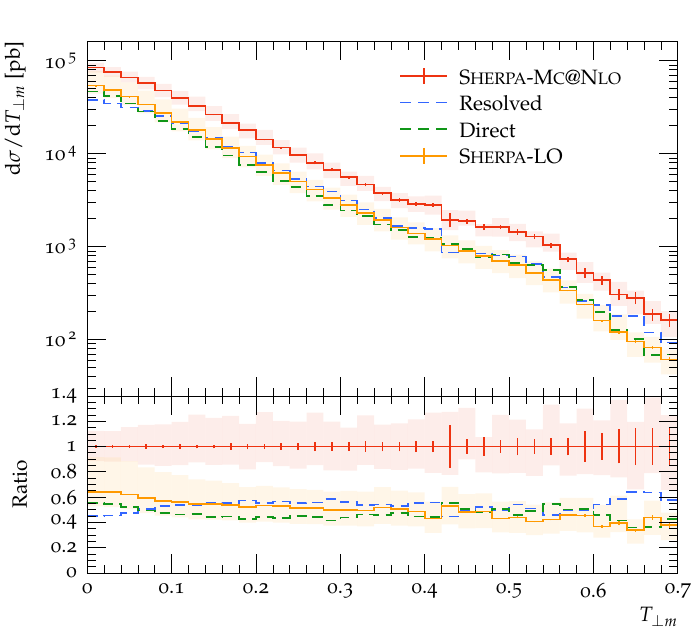}\\
        \includegraphics[width=.4\linewidth]{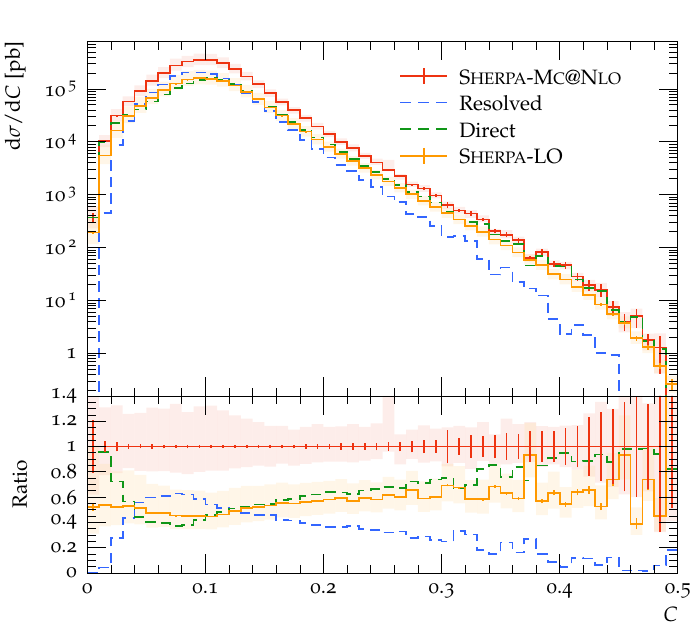} &
        \includegraphics[width=.4\linewidth]{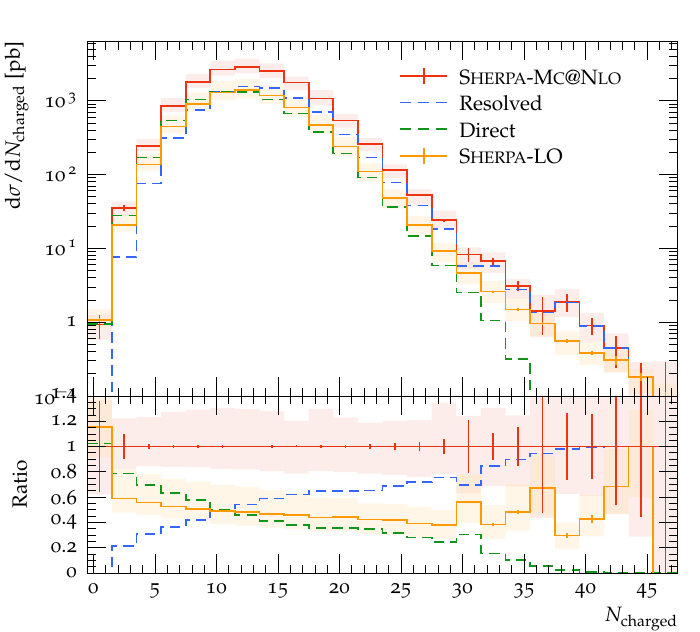}
    \end{tabular}
    \caption{Event shapes observables:
    Transverse thrust major (upper left) and minor (upper right) distributions, $C$ parameter (lower left) and charged particle multiplicity distribution $N_\textrm{charged}$ inside the detector acceptance of $\left| \eta \right| < 4$ (lower right), all obtained with
    the \protect\Sherpa simulation at \protect\MCatNLO accuracy, compared with results at LO.}
    \label{fig:eventshapes}
\end{figure}
Moving on to event shapes, we study distributions of transverse thrust and transverse thrust minor in the upper panel of Fig.\ ~\ref{fig:eventshapes}. 
In both observables, the direct and the resolved component contribute to approximately equal amounts throughout the whole parameter space. 
This is in contrast to the $C$-parameter distribution, depicted in Fig.\ ~\ref{fig:eventshapes} (lower left panel), where we can identify distinct regions where one of the two components dominate. 
While the direct component contributes mostly at the lowest value and in the tail of the distribution, the resolved component contribution reaches up to about 60\% near the global maximum of $C\approx 0.1$. 
Looking at the hadron multiplicity distribution in Fig.\ ~\ref{fig:eventshapes} (lower right), we see -- in agreement with expectations -- that the low-multiplicity states are dominantly produced by direct photo-production, while the high-multiplicity tail is determined by the resolved component. 
It is worth noting here that the effect of multiple-parton interactions is generally mostly negligible at the \EIC collision energies; it does, however play a significant role in high-multiplicity states, therefore still necessitating a tuning of the modelling.

\begin{figure}[!ht]
    \centering
    \begin{tabular}{cc}
        \includegraphics[width=.4\linewidth]{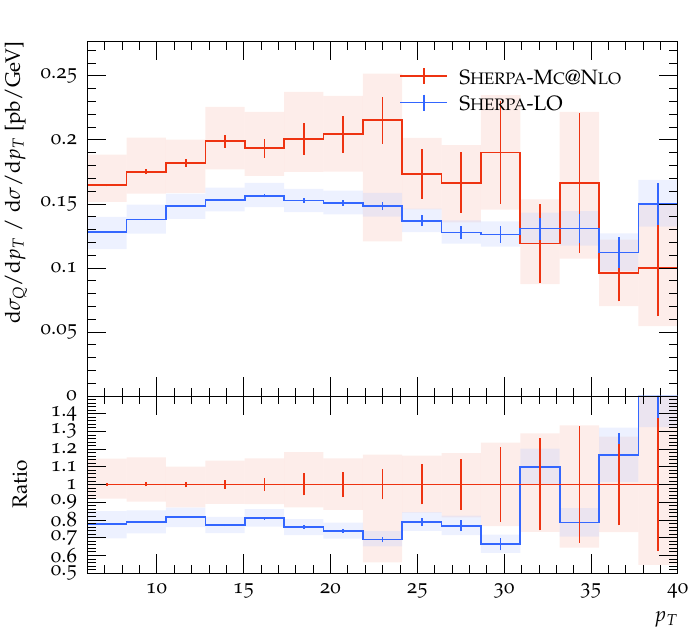} &
        \includegraphics[width=.4\linewidth]{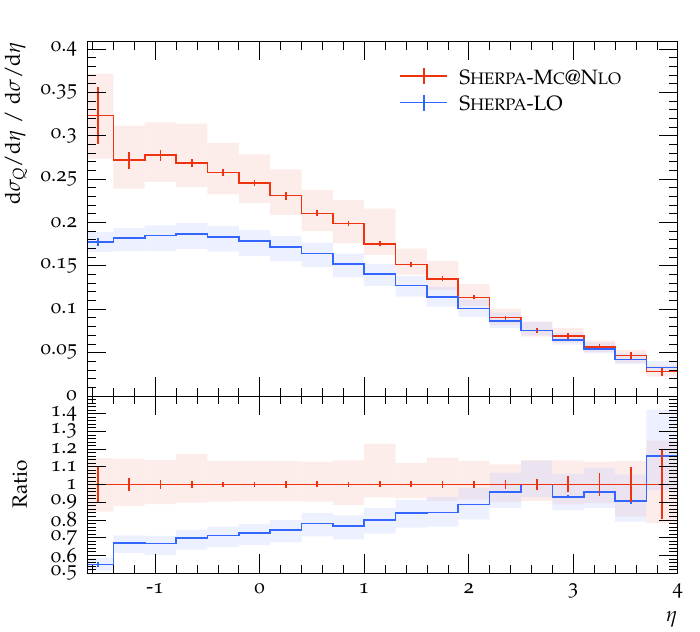}
    \end{tabular}
    \caption{Distributions of heavy quark jet transverse momentum $p_T$ (left) and pseudo-rapidity $\eta$ (right) of the \protect\Sherpa simulation with \protect\MCatNLO accuracy, compared with results at LO.}
    \label{fig:heavy-jet-ratios}
\end{figure}
As a last pair of observables, we look at the ratio of $c$- and $b$-quark jets to light-quark jets in transverse momentum and pseudo-rapidity distributions in Fig.\ ~\ref{fig:heavy-jet-ratios}. 
Heavy-quark jets are almost exclusively produced through the direct process and in the region $\eta<0$, where they make up almost a third of the overall activity. 
As expected, the heavy-quark jets are predominantly $c$-quark jets and their tagging will benefit from a cut on the $x_\gamma$ value. 

\begin{figure}[h!]
    \centering
    \includegraphics[width=.6\linewidth]{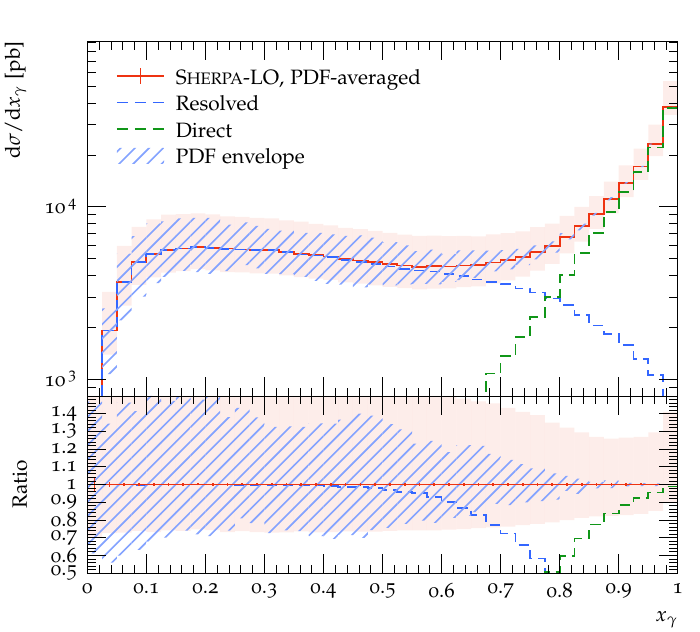}
    \caption{Distributions of $x_\gamma$ of the \protect\Sherpa simulation at parton-level with LO accuracy and without MPI effects, averaged over 11 photon PDF sets. The PDF envelope shows the per-bin variation among the PDF sets. }
    \label{fig:pdf-var}
\end{figure}
As indicated above, the $x_\gamma$ distribution is particularly important for tagging and discerning different event types, and it also acts as an excellent discriminator of resolved and unresolved photon interactions.
Given the relative cross sections, it is also clearly important to improve the theoretical predictions obtained in resolved photo-production. 
However, at the moment, this domain suffers from large uncertainties due to the somewhat outdated photon PDFs, obtained about 2 decades earlier, and with proton PDFs from the same period for the description of photo-production at \HERA. 
This is further highlighted in Fig.\ ~\ref{fig:pdf-var}, where we generated one million parton-level events at LO for 11 photon PDFs sets each, including a 7-point scale variation, using the same settings as before. 
Creating an envelope from the minimal/maximal values per bin among the different PDF sets, we find that the variations are of similar size as the scale uncertainties and reach up to 50\% of the binned cross-section. 
This finding also suggests that a renewed fitting exercise of photon PDFs is of great importance for the upcoming EIC and the full exploitation and understanding of its data.
\section{Summary}\label{Sec:summary}

In this paper we presented the first fully differential hadron-level study at Next-to-Leading order accuracy for the photo-production of inclusive QCD final states at the \EIC.
For our predictions, obtained with \Sherpa, we assumed beam energies of 18 (275) GeV for the electron (proton) beam and analysed the events with inclusive-jet as well as event-shape and heavy-jet observables. 
Photo-production represents an interesting laboratory to study a large variety of sometimes subtle QCD effects, by providing access to a wide range of exclusive and inclusive production channels.
Our results will also be instrumental for renewed efforts to fit PDFs for the photon in collinear factorisation.

We identified the photon PDFs as probably the most obvious bottleneck in precise photo-production phenomenology for the \EIC, as they are based on fits from 2 decades ago, using contemporary proton PDFs and simulation tools of that period. 
An updated fit based on modern methodology including error estimates is urgently needed: a coherent and qualitatively satisfying theoretical description of resolved photo-production necessitates a PDF fit based on modern proton PDFs, with consistent settings, for example for $\alpha_S$. 

This is an urgently needed input to arrive at state-of-the-art baseline predictions for inclusive QCD final-state production in its dominant channel and it will ultimately provide us with a theoretical sound baseline for inclusive QCD studies at the \EIC.

\section*{Acknowledgements}
We are grateful for enlightening conversations with Paul Newman.
We are indebted to our colleagues in the \Sherpa collaboration, for numerous discussions and technical support.
F.K.\ gratefully acknowledges funding as Royal Society Wolfson Research fellow. 
P.M.\ is supported by the STFC under grant agreement ST/P006744/1.

\bibliographystyle{unsrt}
\bibliography{journal,analyses}
\end{document}